\documentclass[prl,showpacs,twocolumn]{revtex4}

\usepackage{graphicx}

\begin{document}

\title{
Effects of superconducting gap anisotropy on the flux flow resistivity in ${\bf Y(Ni_{1-x}Pt_x)_2B_2C}$
}
\author{K. Takaki}
\thanks{Present address : Yokohama R\&D Laboratories, The Furukawa Electric Co., LTD., 2-4-3, Okano, Nishi-ku, Yokohama 220-0073, Japan}
\author{A. Koizumi}
\author{T. Hanaguri}
\thanks{Corresponding author}
\author{M. Nohara}
\author{H. Takagi}
\author{K. Kitazawa}
\affiliation{
Department of Advanced Materials Science, The University of Tokyo,\\
7-3-1, Hongo, Bunkyo-ku, Tokyo 113-0033, Japan
}
\author{Y. Kato}
\author{Y. Tsuchiya}
\thanks{Present address : Research Center for Quantum Effect Electronics, Tokyo Institute of Technology, 2-12-1, O-okayama, Meguro-ku, Tokyo 152-8552, Japan}
\author{H. Kitano}
\author{A. Maeda}
\affiliation{
Department of Basic Science, The University of Tokyo,\\
3-8-1, Komaba, Meguro-ku, Tokyo 153-8902, Japan
}

\begin{abstract}
The microwave complex surface impedance $Z_s$ of ${\rm Y(Ni_{1-x}Pt_x)_2B_2C}$ was measured at 0.5~K under magnetic fields $H$ up to 7~T.
In nominally pure ${\rm YNi_2B_2C}$, which is a strongly anisotropic $s$-wave superconductor, the flux flow resistivity $\rho_f$ calculated from $Z_s$ was twice as large as that expected from the conventional normal-state vortex core model.
In Pt-doped samples where the gap anisotropy is smeared out, the enhancement of $\rho_f$ is reduced and $\rho_f$ approaches to the conventional behavior.
These results indicate that energy dissipation in the vortex core is strongly affected by the anisotropy of the superconducting gap.
\end{abstract}

\pacs{74.60.Ec, 74.25.Jb, 74.25.Nf, 74.70.Dd}
\maketitle

In superconductors with anisotropic superconducting gap (SG), low-energy quasi-particle (QP) excitations near the gap nodes play an important role.
Volovik examined the effects of gap nodes on the QP density of states (DOS) in the mixed state~\cite{Volovik}.
He pointed out that the QP DOS at the Fermi energy $\epsilon_F$ in superconductors with line nodes varies proportionally to $\sqrt{H}$, in contrast to the case of conventional superconductors with an isotropic SG where the QP DOS at $\epsilon_F$ is linear in $H$, that is, proportional to the number of vortices.
Such an enhancement of the QP DOS in anisotropic superconductors has been observed experimentally in $d$-wave high-$T_c$ cuprates~\cite{Moler,Nohara} and the heavy fermion superconductor ${\rm UPt_3}$~\cite{Ramirez} by electronic specific heat measurements.

Besides the QP DOS, gap nodes are expected to affect the flux flow resistivity $\rho_f$.
Basically, $\rho_f$ is a measure of the QP relaxation {\it in and around} the vortex core and can provide different information from the QP DOS.
In dirty superconductors where the QP mean free path $l$ is shorter than the coherence length $\xi$, the vortex core can be treated as a bulk normal metal with a radius of the order of $\xi$~\cite{Bardeen}.
As a result, $\rho_f$ is simply given by $\rho_f\sim\rho_nH/H_{c2}$ regardless of the anisotropy of the SG.
In moderately clean superconductors where $\xi<l<\xi\epsilon_F/\Delta$ ($\Delta$; gap amplitude), however, the vortex core cannot be regarded as a simple normal metal at all, since the QPs are reflected at the edge of the vortex core before they experience impurity scattering.
In this regime, $\rho_f$ is given by $\rho_f=H/(ne<\omega_0\tau_c>_F)$~\cite{Kopnin}.
Here $n$ is a carrier density, $e$ is a unit charge, $\omega_0\sim \Delta^2/\epsilon_F$ is the precession frequency of the lowest Andreev bound state~\cite{Caroli}, $\tau_c$ is the relaxation time in the vortex core and $<...>_F$ denotes the average over the Fermi surface.
In such a case, nodes in the SG should affect both of $\omega_0$ and $\tau_c$.
First, $\omega_0$ in anisotropic superconductors depends on the QP momentum and should be very small near the gap nodes.
Second, $\tau_c$ can be short because the QP DOS which contributes to the impurity scattering is enhanced in anisotropic superconductors.
Both of these factors are expected to enhance $\rho_f$ from that of clean isotropic superconductors.

To date, ${\rm UPt_3}$ is the only clean anisotropic superconductor in which $\rho_f$ has been measured.
The enhancement of $\rho_f$ has actually been observed~\cite{Kambe,Luetke}.
However, since ${\rm UPt_3}$ is a complicated odd-parity superconductor~\cite{Tou}, it should be noted that many factors other than gap anisotropy can contribute to $\rho_f$.
In fact, L\"utke-Entrup {\it et al.} attributed the origin of the $\rho_f$ enhancement to the unconventional vortex core structure which is brought by the two-component order parameter~\cite{Luetke}.
In addition, Kato showed that the parity and chirality of the pair wave function affect $\tau_c$ in the moderately clean regime~\cite{Kato}.
Therefore, it is still unclear whether $\rho_f$ enhancement in ${\rm UPt_3}$ is brought by the gap anisotropy or by the other unconventional nature of superconductivity.

To make clear the effects of gap anisotropy on $\rho_f$, measurement on a {\it clean, anisotropic and $s$-wave} superconductor is indispensable.
From this point of view, we measured the $\rho_f$ of ${\rm Y(Ni_{1-x}Pt_x)_2B_2C}$.
The temperature and magnetic-field dependence of the electronic specific heat~\cite{Nohara2,Izawa}, thermal conductivity~\cite{Boaknin,Izawa2}, and NMR relaxation rate~\cite{Zheng} indicate that there are low-lying QP states near $\epsilon_F$ even in the superconducting state.
This is strong evidence for the huge anisotropy of the SG in ${\rm YNi_2B_2C}$.
Recently, Izawa {\it et al.} suggest from thermal conductivity measurements under various magnetic field directions that there are point-like gap nodes along the [100] and [010]-directions~\cite{Izawa2}.
In addition, it is reported that the introduction of impurities {\it reduces} the low-lying QP states and opens the gap all over the Fermi surface with little effect on $T_c$~\cite{Nohara2,Yokoya}.
Such behavior is totally different from that expected in $d$- or $p$-wave superconductors~\cite{Borkowski_Fehrenbacher} and provides a strong evidence of {\it anisotropic $s$-wave} superconductivity in ${\rm YNi_2B_2C}$.
Therefore, by comparing the results in pure (anisotropic) and impurity-doped (isotropic) ${\rm YNi_2B_2C}$, the effects of SG anisotropy on $\rho_f$ can be examined without the complication of other effects.

\begin{table}[b]
\caption{
Material parameters of the ${\rm Y(Ni_{1-x}Pt_x)_2B_2C}$ crystals studied.
Here, $\rho_{res}$ is the residual resistivity measured just above $T_c$ and $\delta$ is the skin depth at 44~GHz.
The QP mean free path $l$ was estimated from $\rho_{res}$ and a free electron model using a carrier density $n=3\times10^{22}$~cm$^{-3}$.
}
\begin{tabular}{ccccccc}
x & $T_c$ (K) & $H_{c2}$(0) (T) & $\rho_{res}$ ($\mu\Omega$cm) & $\xi$ (\AA) & $l$ (\AA) & $\delta$ (\AA) \\
\hline
0 & 15.4 & 8.0 & 0.87 & 65 & 1500 & 2200 \\
0.05 & 13.7 & 5.0 & 20.5 & 80 & 63 & 11000 \\
0.2 & 12.1 & 4.3 & 35.5 & 90 & 38 & 14000 \\
\end{tabular}
\end{table}

Samples were cut from the same single crystal boules used in previous studies~\cite{Nohara2,Yokoya}.
As an impurity, Pt was used instead of Ni.
The relevant material parameters are listed in table I.
The sample with x=0 is in the moderately clean regime and the impurity-doped samples are in the dirty regime.
In order to avoid pinning-related effects, $\rho_f$ should be measured at high enough currents or at frequencies higher than the pinning frequency $f_p$~\cite{Gittleman}.
We adopted the latter method to avoid the self-heating of the sample.
In nominally pure ${\rm YNi_2B_2C}$, $f_p$ is estimated to be below 3~MHz~\cite{Oxx}.
In other conventional superconductors, $f_p$ is at most of the order of 10$^2$ MHz~\cite{Gittleman}.
Consequently, pinning-related effects can be avoided completely at microwave frequencies.
However, direct resistivity measurement is difficult at microwave frequencies because of the skin effect.
Therefore, we measured the surface impedance $Z_s=R_s+iX_s$ using a  cavity perturbation method and calculated the real part of the resistivity $\rho_1$ using the relation $\rho_1=2R_sX_s/\mu\omega$, where $\mu$ is the permeability and $\omega$ is the angular frequency.
Samples were located at the center of a cylindrical Cu cavity which resonate at 44~GHz in the TE$_{011}$ mode.
Microwave magnetic fields were applied along the $c$-axis of the crystal to measure the in-plane response.
The surface resistance $R_s$ and the surface reactance $X_s$ were obtained from the changes in the quality factor and the resonance frequency of the cavity, respectively~\cite{Klein}.
The absolute value of $Z_s$ was determined by comparison with the dc resistivity $\rho$ above $T_c$ assuming the Hagen-Rubens relation $R_s=X_s=\sqrt{\mu\omega\rho/2}$.
This relation can be applied in the local regime where $l$ is much shorter than the skin depth $\delta=\sqrt{2\rho/\mu\omega}$.
Strictly speaking, the sample with x=0 is not deep in the local regime since $l$ and $\delta$ were estimated to be of the same order of magnitude.
In the non-local regime, $X_s$ becomes larger than $R_s$ and both $R_s$ and $X_s$ are larger than the local value~\cite{Abrikosov}.
Therefore, the absolute value of $Z_s$ for the sample with x=0 could be somewhat underestimated.
In the present study, the absolute value is not crucial to the final conclusion.
Samples were cooled down to 0.5~K by a laboratory-made $^3$He refrigerator.
In all the measurements, static magnetic fields were applied along the $c$-axis of the crystals.

The magnetic field dependence of the surface impedance at 0.5~K for each sample is plotted as a function of $\sqrt{H/H_{c2}}$ in Fig.~1.
In the absence of magnetic fields, the response is purely reactive with $R_s\sim 0$ and $X_s=\mu\omega\lambda_0$, where $\lambda_0$ is the penetration depth in the Meissner state.
We estimate $\lambda_0$ to be 500~${\rm \AA}$ for the sample with x=0.
This value is similar to $\lambda_0$ obtained by the previous microwave measurement~\cite{Izawa}, but is rather short compared to the other estimates~\cite{Rathnayaka_Cywinski}.
Part of this difference may come from the non-locality of the response.

\begin{figure}[b]
\includegraphics[angle=0,scale=0.5]{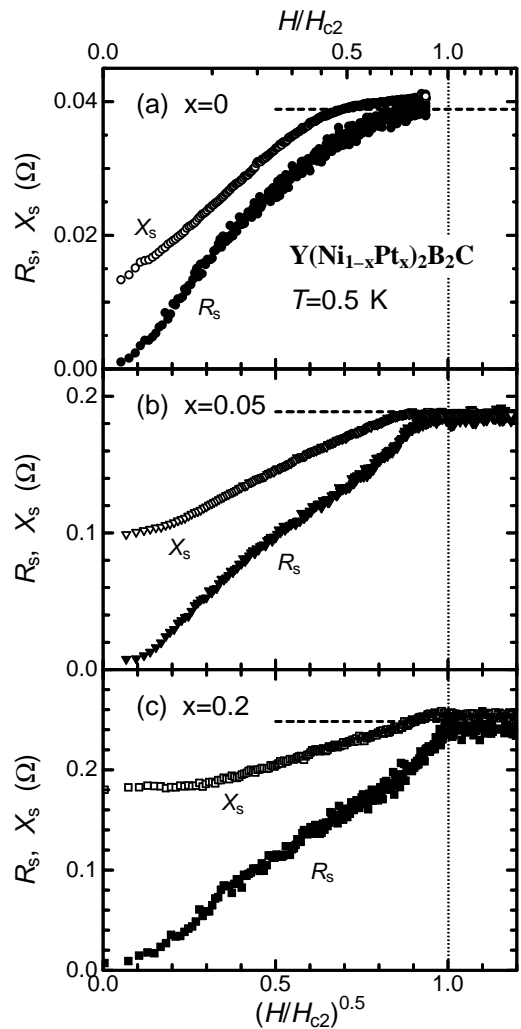}
\caption{
Magnetic field dependence of the surface impedance of ${\rm Y(Ni_{1-x}Pt_x)_2B_2C}$ plotted as a function of $\sqrt{H/H_{c2}}$.
Horizontal dashed lines denote the normal state values just above $T_c$.
}
\end{figure}

In the mixed state, $R_s$ and $X_s$ increase with increasing $H$.
At high fields, both $R_s$ and $X_s$ saturate to the value just above $T_c$, which is given by $\sqrt{\mu\omega\rho_{res}/2}$, where $\rho_{res}$ is the resistivity just above $T_c$.
In the sample with x=0.2, this saturation occurs just at $H_{c2}$ with a clear kink, as expected.
However, with decreasing impurity content or increasing SG anisotropy, both $R_s$ and $X_s$ become to saturate even below $H_{c2}$ and the kink structure is broadened.
This behavior is not caused by the inhomogeneity of the sample, since the broadening is most apparent in the cleanest sample.
Therefore, we can conclude that, in clean ${\rm YNi_2B_2C}$, there is a distinct field region below $H_{c2}$ where the QP response is almost as same as that in the normal state.
In the sample with x=0, this region extends to about half of $H_{c2}$.
This behavior may be related to the de Haas-van Alphen effect in the mixed state~\cite{Terashima} and the large DOS at $\epsilon_F$ in the mixed state observed by scanning tunneling spectroscopy~\cite{Sakata}.

Next we try to extract $\rho_f$ from $R_s$ and $X_s$.
In general, $Z_s$ in the mixed state consists of several contributions and $\rho_1$, which is directly calculated from $Z_s$, is different from $\rho_f$.
Since our measurements were performed at low enough temperatures ($T/T_c<0.04$), and high enough frequencies ($\omega>>\omega_p$), contributions from thermally activated QPs and the effects of vortex pinning can be safely neglected.
One of the remaining potential contributions is the effect of flux flow.
Another important factor that we should consider is the contribution from delocalized QPs.
In a superconductor with an isotropic SG, all the QPs in the mixed state are localized in the vortex core.
On the other hand, in a strongly anisotropic superconductor, such as nominally pure ${\rm YNi_2B_2C}$, QPs are generated not only in the vortex core, but also {\it outside} of the core by the Doppler energy shift from the supercurrent around the vortex~\cite{Volovik}.
The existence of such delocalized QPs has been confirmed in ${\rm YNi_2B_2C}$ by electronic specific heat measurements~\cite{Nohara2,Izawa} and in ${\rm LuNi_2B_2C}$ by thermal conductivity measurements~\cite{Boaknin}.
To analyze $Z_s$ in the mixed state under the influence of the delocalized QPs, we use the formulation of $Z_s$ given by Coffey and Clem~\cite{Coffey}.
If we neglect the effect of vortex pinning, $Z_s$ is given by,

\begin{equation}
\label{Zs}
Z_s=i\mu\omega\lambda_0\sqrt{\frac{(1-bs)-i(b+s)}{1+s^2}}
\end{equation}

\noindent
where $b=\rho_f/\mu\omega\lambda_0^2$, $s=\mu\omega\lambda_0^2\sigma_{nf}$ and $\sigma_{nf}$ is the conductivity of the delocalized QPs.
Here, $b$ and $s$ are measures of the contributions from the flux flow and the delocalized QPs, respectively.
If we set $s=0$, Eq.~\ref{Zs} is reduced to the pure flux flow state without delocalized QPs and $b=0$ denotes the usual two-fluid response without vortices.
In the latter case, all the QPs are delocalized as in the case of thermally activated QPs and $s=\omega\tau_{nf} f_{nf}/(1-f_{nf})$, where $\tau_{nf}$ is the relaxation time of the delocalized QPs and $f_{nf}$ is the normal fluid fraction.

To judge which contribution is dominant in ${\rm Y(Ni_{1-x}Pt_x)_2B_2C}$, it is useful to introduce the ``impedance plane plot'', where $X_s/X_s(0)$ is plotted against $R_s/X_s(0)$~\cite{Tsuchiya}.
Here $X_s(0)=\mu\omega\lambda_0$ is a surface reactance in the absence of magnetic field.
This plot is sensitive to the underlying mechanism that dominates $Z_s$.
Figure~2 shows the impedance plane plots for ${\rm Y(Ni_{1-x}Pt_x)_2B_2C}$.
The solid and dashed curves correspond to two limiting cases of Eq.~\ref{Zs}, namely, a pure flux flow state ($s=0$) and two-fluid response ($b=0$), respectively.
As is evident from Fig.~2, the behavior of $Z_s$ is reproduced quite well by the assumption of pure flux flow.
Small deviation can be seen at high fields where $R_s$ and $X_s$ are saturated.
Consequently, the contribution of delocalized QPs to $Z_s$ can be neglected in the field range where $R_s$ and $X_s$ is field dependent.
In other words, in this field range, the real part of the resistivity $\rho_1$, calculated from $Z_s$, can be regarded as $\rho_f$ itself.

\begin{figure}[b]
\includegraphics[angle=0,scale=0.4]{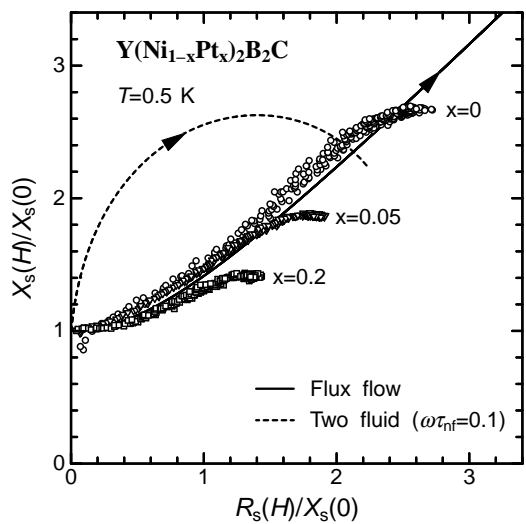}
\caption{
Impedance plane plot of the surface impedance of ${\rm Y(Ni_{1-x}Pt_x)_2B_2C}$.
Solid and dashed lines denote the calculated curves in cases of pure flux flow and the two-fluid response, respectively.
For the calculation of two-fluid model, we used $\omega\tau_{nf}=0.1$.
(See text.)
Arrows indicate the direction when $H$ is increased.
}
\end{figure}

\begin{figure}[t]
\includegraphics[angle=0,scale=0.4]{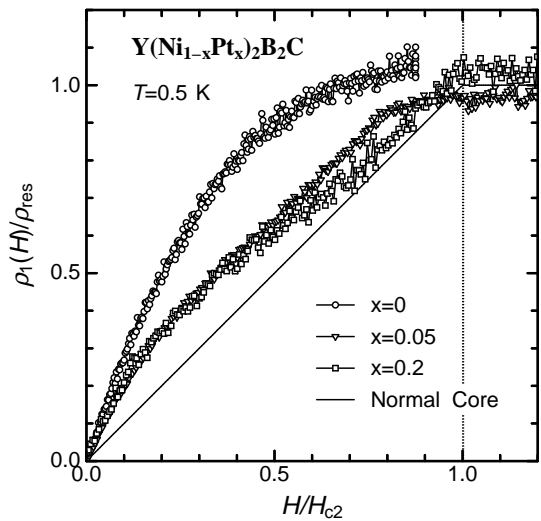}
\caption{
Real part of the complex resistivity as a function of magnetic field.
Vertical and horizontal axes are normalized by $\rho_{res}$ and $H_{c2}$, respectively.
}
\end{figure}

Figure.~3 plots $\rho_1/\rho_{res}$ as a function of $H/H_{c2}$.
In the sample with x=0, in which the SG is strongly anisotropic, $\rho_1$, namely $\rho_f$, below about a half of $H_{c2}$ is almost proportional to $H$ as in the case of the usual flux flow.
However, its slope is twice as large as that expected from the conventional normal core model~\cite{Bardeen}.
We note here that this enhancement is not an artifact caused by the assumption of local electrodynamics.
As described above, the absolute value of $Z_s$ may be somewhat underestimated.
Therefore, correction for non-local behavior further enhances the anomaly.
The enhancement of $\rho_f$ implies that the energy dissipation in the vortex core is strongly enhanced in ${\rm YNi_2B_2C}$.
In impurity-doped crystals where the anisotropy of the SG is smeared out, this enhancement is reduced except at very low fields and $\rho_f$ approaches to the behavior expected in the normal core model~\cite{Bardeen}.
Although the origin of the residual enhancement of $\rho_f$ in impurity-doped crystals is unclear now, the effect of gap anisotropy on $\rho_f$ is evident.

The observed enhancement factor of 2 is similar to that of ${\rm UPt_3}$ (1.6 $\sim$ 4.7)~\cite{Kambe,Luetke}.
However, unlike in the case of ${\rm UPt_3}$, we can safely conclude that the enhancement of $\rho_f$ in ${\rm YNi_2B_2C}$ is solely brought by the SG anisotropy, since ${\rm YNi_2B_2C}$ is a ``simple'' superconductor except for the huge anisotropy of the SG.
As mentioned before, $\rho_f$ is affected by the SG anisotropy through $\omega_0$ and/or $\tau_c$.
Future work may examine which path is dominant.
It should be noted that such an enhancement of $\rho_f$ can not be expected in a moderately clean {\it isotropic} superconductor, since $\omega_0\tau_c$ is always large over the Fermi surface in this case.

Finally, we briefly comment on the absence of the contribution from delocalized QPs to $Z_s$.
If the parameter $s$ is small compared to unity, the effect of delocalized QPs on $Z_s$ is small.
To evaluate $s$ in the present experiment, we assume that $\tau_{nf}$ in the mixed state is not so different from that in the normal state.
If that is the case, $\tau_{nf}$ can be estimated from $\rho_{res}$.
In the sample with x=0, $\omega\tau_{nf}$ is estimated to be of the order of 0.1 and is much smaller in the impurity-doped samples.
This means that even if the normal fluid fraction $f_{nf}$ becomes 0.5, $s=\omega\tau_{nf} f_{nf}/(1-f_{nf})$ is only 0.1 or less.
In other words, even at microwave frequencies, one can selectively extract information on the flux flow or the dynamics of the QPs in the vortex core from $Z_s$ measurements.
This feature is complementary to that of the thermal conductivity measurement which selectively senses the dynamics of delocalized QPs~\cite{Boaknin}.
Considering that the electronic specific heat measures the total QP DOS, combination of these techniques should be important for investigations of the electronic states of the mixed state.

In summary, we measured $Z_s$ of the anisotropic $s$-wave superconductor ${\rm YNi_2B_2C}$ to study the effects of SG anisotropy on $\rho_f$.
We found that $\rho_f$ in the nominally pure sample, in which the SG has a huge anisotropy, is greatly enhanced from that expected from the conventional normal-state vortex core.
This enhancement was reduced in impurity-doped samples where the SG anisotropy is smeared out.
Such behavior indicates that the quasi-particle relaxation in and around the vortex core is sensitive to the superconducting gap anisotropy.
These results imply that $\rho_f$ measurement provides useful information for the microscopic understanding of the vortex state of anisotropic superconductors.

\begin{acknowledgements}
The authors thank to Y. Matsuda and K. Izawa for helpful discussions.
They also thank D. G. Steel for a critical reading of the manuscript.
This work was supported by SORST, Japan Science and Technology Corporation and the Grant-in-Aid for Scientific Research from the Ministry of Education, Science, Sports and Culture of Japan.
\end{acknowledgements}

\end{document}